\documentclass[12pt]{article}
\usepackage{amsmath,amssymb,epsf,color}
\usepackage{mathrsfs}
\usepackage{amsfonts}

\makeatletter \@addtoreset{equation}{section} \makeatother

\newcommand{\hoch}[1]{$\, ^{#1}$}

\def\ft#1#2{{\textstyle{\frac{\scriptstyle #1}{\scriptstyle #2} } }}

\begin{document}

\begin{flushright}
\hfill{ \ CAS-KITPC/ITP-257\ \ \ \ }
\end{flushright}

\vspace{25pt}
\begin{center}
{\large {Holographic de Sitter Universe}}

\vspace{15pt}

Miao Li\hoch{} and Yi Pang\hoch{}

\vspace{10pt}

\hoch{} {\it Kavli Institute for Theoretical Physics,\\
State Key Laboratory of Frontiers in Theoretical Physics,\\
 Institute of Theoretical Physics, Chinese Academy of Sciences, Beijing 100190, P.R.China}

\vspace{40pt}

\underline{ABSTRACT}

\end{center}

We propose to embed de Sitter space into five dimensional anti-de Sitter space to compute some physical
quantities of interest, using the AdS/CFT correspondence. The static de Sitter can be considered as the
conformal structure of the boundary of the hyperbolic $AdS_5$ with a horizon, thus energy as well as entropy can be
computed. The global dS can be embedded into a half-global $AdS_5$, and the dS entropy can be regarded
as entanglement entropy in this case. Finally, the inflationary dS can also be regarded as the boundary
of $AdS_5$, and this can be extended to include general cosmology with a positive cosmological constant,
however at the price of introducing a naked null singularity in the bulk unless the cosmology is pure de Sitter space.

 \vspace{15pt}

\thispagestyle{empty}




\newpage
\section{Introduction}

The problem of dark energy is one of the most profound problems in fundamental physics and
cosmology. More specifically, we do not yet understand a positive cosmological constant
and physics in de Sitter space.

Understanding de Sitter space can be viewed as a first step toward understanding dark energy
and the origin of the universe with dark energy.
Many attempts have been made to understand the origin of de Sitter space as well as physics
in de Sitter space, such as the string landscape scenario \cite{Susskind:2003kw}-\cite{Kachru:2003aw} ,
matrix model of de Sitter space \cite{Li:2001ky}, the dS/CFT correspondence \cite{Maldacena:1998ih}-\cite{Strominger:2001pn},
and resonant bubble in AdS \cite{Freivogel:2005qh,Alberghi:1999kd,{Lowe:2010np}}\footnote{Some comments on embedding inflation in AdS/CFT can be found in \cite{Lowe:2007ek}}, to name a few. However, no serious
progress has been made by far.

The present work is motivated by our previous study on using metamaterials to mimic de Sitter
space and our calculation of the Casimir energy in de Sitter space \cite{Li:2009pm,Li:2009dw}.

It is known that de Sitter space can be embedded into anti-de Sitter space as its boundary,
while some aspects of quantum field theory on de Sitter space based on the
AdS/CFT correspondence were explored in \cite{Koyama:2001rf}-\cite{Maldacena:2010un} and references therein. We carry out a deeper study in this paper. The static de Sitter can be considered as the conformal structure of the boundary of the hyperbolic $AdS_5$ with a horizon, thus energy as well as entropy can be computed. Of course, the volume of the horizon of the hyperbolic AdS is infinite, we need to
introduce an infrared cut-off. Interestingly, this cut-off corresponds to a cut-off separating a stretched
horizon from the real horizon in the static de Sitter, the conformal boundary of the hyperbolic AdS.
As we shall see, energy computed using the AdS/CFT picture is proportional to the radius of dS when
the infrared cut-off is interpreted in the above fashion. The energy/entropy ratio computed has
the same functional form as we expect from the dS physics.

The global dS can be embedded into a half-global $AdS_5$, and the dS entropy can be regarded
as entanglement entropy in this case. Finally, the inflationary dS can also be regarded as the boundary
of $AdS_5$, the extension of this includes general cosmology with a positive cosmological constant. However, an
imperfect point is that this embedding has null singularity for a general cosmology.

In the next section, we embed the static de Sitter space as well as the global de Sitter space
into $AdS_5$, and we embed the inflationary de Sitter space into $AdS_5$ in sect.3 and we show that
this embedding can be generalized to a general cosmology with a positive cosmological constant.

\section{Holographic Static de Sitter}

Both energy and entropy are well-defined in the static patch of de Sitter space:
\begin{equation}\label{desitter}
    ds^2_4=-(1-\frac{r^2}{\ell^2})dt^2+(1-\frac{r^2}{\ell^2})^{-1}dr^2+r^2d\Omega^2_2,
\end{equation}
Thermodynamics was considered by Gibbons and Hawking in  \cite{Gibbons:1977mu}, the de Sitter space entropy is still
proportional to the area of the horizon. By embedding de Sitter space to five dimensional anti-de Sitter space, we
shall show that this entropy can be explained as the usual black hole entropy in the hyperbolic patch of 5D anti-de Sitter
space, as well as the entanglement entropy of the two halves of the boundary of a half-global (time-dependent ) anti-de
Sitter space.

We first consider embedding the static de Sitter to 5D anti-de Sitter. Since only the conformal structure is relevant
when de Sitter space is taken as the boundary of anti-de Sitter, we perform the following transformation.
By multiplying a conformal factor $\Omega^2=(1-\frac{r^2}{\ell^2})^{-1}$, above metric can be mapped to the metric on $R\times H^3$
\begin{equation}\label{hypobolic}
    ds^2_4=-dt^2+\frac{d\rho^2}{1+\rho^2/\ell^2}+\rho^2d\Omega^2_2,
\end{equation}
where $r$ is related to $\rho$ by the following transformation
\begin{equation}
    r=\frac{\rho}{\sqrt{1+\rho^2/\ell^2}}.
\end{equation}
From this expression, it is straightforward to see that the de Sitter horizon is mapped to $\rho=\infty$. But usually,
when discussing physics in a space time with a nondegenerate horizon, the boundary condition is always put at the
stretched horizon which is away from the real horizon by a proper distance in Planck unit. In terms of coordinate $r$,
the stretched horizon is at
\begin{equation}
    r_s\simeq\ell-\ell_p^2/2\ell,
\end{equation}
where $\ell_p$ is the proper distance between the stretched horizon and the real horizon, and can be considered
to be on the scale of the Planck length. $r_s$ is mapped to an infrared cut-off of $\rho$, namely the maximum value of $\rho$
\begin{equation}
    \rho_{\rm{max}}\simeq\ell^2/\ell_p.
\end{equation}
To a field conformally coupled to gravity, the background metric (\ref{desitter}) and (\ref{hypobolic}) are equivalent up to some quantum mechanical anomaly. However, as we will see the anomaly contribution to physical quantities of interest of us is negligible.

Before proposing  the holographic dual of static de Sitter, we
should mention that it is known for a long time  that to an observer
sitting at the center of static de Sitter, the global de Sitter
vacuum of a quantum field appears to be a thermal state with
temperature $1/(2\pi\ell)$ \cite{Gibbons:1977mu}. To keep this
feature in the conformal metric (\ref{hypobolic}), we require the
global de Sitter vacuum be described by a thermal state on $R\times
H^3$ with temperature $1/(2\pi\ell)$.

Now we propose the holographic dual to the static de Sitter is the
the hyperbolic foliatation of pure $AdS_5$
\begin{equation}\label{hyperbolic}
    ds^2_5=-(\frac{r^2}{\ell^2}-1)dt^2+(\frac{r^2}{\ell^2}-1)^{-1}dr^2+\frac{r^2}{\ell^2}dH_3^2\,,\qquad(r\geq\ell)
\end{equation}
with
\begin{equation}\label{}
dH_3^2=
 (1+\frac{\rho^{2}}{\ell^{2}})^{-1}d\rho^{2} +
 \rho^{2}d\Omega^{2}_2,
\end{equation}
the boundary of (\ref{hyperbolic}) is conformal to the metric in
(\ref{hypobolic}), thus is conformal to the static de Sitter. The
hyperbolic $AdS_5$ has a horizon thus has a temperature
$1/(2\pi\ell)$ coinciding with the temperature of boundary field
theory on $R\times H^3$. Therefore, this geometry provides a
holographic description of boundary thermal state with temperature
$1/(2\pi\ell)$ which is the global de Sitter vacuum of quantum field
observed by the center observer. We now compute the energy and
entropy of this state \textsl{via} AdS/CFT method
\cite{Balasubramanian:1999re}-\cite{Emparan:1999gf}. They are given
by
\begin{equation}\label{energy}
    E=\frac{3V_{H^3}}{64\pi G_5\ell},
\end{equation}
\begin{equation}\label{entropy}
    S=\frac{V_{H^3}}{4G_5}.
\end{equation}
In above expressions, $V_{H^3}$ is the volume of hyperbolic space
\begin{equation}\label{}
    V_{H^3}=\Omega_2\int^{\rho_{\rm{max}}}_0\frac{\rho^2d\rho}{\sqrt{1+\rho^2/\ell^2}}
    \simeq\ft{1}{2}\Omega_2l^3\rho^2_{\rm{max}}=2\pi\ell^5/\ell_p^2.
\end{equation}
Inserting this value into (\ref{energy}) and (\ref{entropy}), we obtain
\begin{equation}\label{energy}
    E=\frac{3\ell^3}{32 G_5}\frac{\ell}{\ell_p^2},
\end{equation}
\begin{equation}
    S=\frac{\pi\ell^3}{2 G_5}\frac{\ell^2}{\ell_p^2}.
\end{equation}
 the dimensionless quantity $\ell^3/G_5$ should to be replaced by a certain pure number associated with the degrees of freedom in the
 dual field theory. For instance, in the well-known AdS/SYM correspondence,
  \begin{equation}
   \pi\ell^3/2G_5\rightarrow N^2.
 \end{equation}

  Thus we find that in the de Sitter horizon, energy of the global de Sitter vacuum of field observed by the center observer is proportional to
  the size of de Sitter horizon as predicted in \cite{Candelas:1978gf,Li:2009pm,Li:2009dw} and the entropy is proportional to the
  area of horizon previously found in \cite{Dowker:1994fi} as geometric or entanglement entropy of field in static de Sitter.
  Based on the formulas of energy and entropy, one can compute the energy per degree of freedom. It is given by
  \begin{equation}\label{ratio1}
   E/S=\frac{3}{16\pi \ell}.
  \end{equation}
 This ratio, unlike the total energy and entropy, is independent of
 the two undetermined parameters $\ell^3/G_5$ and $\ell_p^2$.

 While for the pure de Sitter vacuum, the energy momentum tensor of cosmological constant is $T^{\mu}_{\nu}=-3\delta^{\mu}_{\nu}/(8\pi G_4\ell^2)$.
 The corresponding energy inside the horizon can be computed as
 \begin{equation}
    E_{\Lambda}=\int^{\ell}_0T^{\mu}_{\nu}\xi^{\nu} n_{\mu}\sqrt{\gamma}drd\theta d\phi=\frac{\ell}{2G_4},
 \end{equation}
 where $\xi^{\nu}$ is the time Killing vector in the horizon and $ n^{\mu}$ is the future directed unit normal vector of constant time surface. Since the Gibbons-Hawking entropy of de Sitter is
 \begin{equation}
    S_{\Lambda}=\frac{\pi\ell^2}{G_4},
 \end{equation}
 thus energy per degree of freedom of pure de Sitter is given by
 \begin{equation}\label{ratio2}
    E_{\Lambda}/S_{\Lambda}=\frac{1}{2\pi\ell}.
 \end{equation}
 Comparing (\ref{ratio1}) and (\ref{ratio2}), we find the energy per degree of freedom of the global de Sitter
 vacuum of field observed by the center observer is smaller than that of pure de Sitter by a factor $\frac{3}{8}$.
It is interesting to note that ratio (\ref{ratio2}) is just the de
Sitter temperature. We do not know how to explain the mismatch
factor $3/8$ in (\ref{ratio1}), nevertheless, it is apparent that by
taking $dS_4$ as the boundary of the hyperbolic $AdS_5$, there is no
gravity in $dS_4$, thus physical quantities are not expected
to be identical to those in a $dS_4$ with gravity.

In the following, we will show that the entropy (\ref{entropy})
indeed has a holographic interpretation of entanglement entropy
\cite{Myers:2010tj}. To see this clearly, we study $AdS_5$ in
the following coordinates
\begin{equation}\label{global}
    ds^2_5=(1+\frac{r^2}{\ell^2})^{-1}dr^2+r^2(-dt^2/\ell^2+\cosh^2(t/\ell)d\Omega_3^2).
\end{equation}
The conformal boundary of above metric is the global de Sitter space. This metric has the following flat embedding in the $R^{(2,4)}$ as
\begin{equation}
    ds_6^2=-dX_0^2-dX_1^2+dX^2_2+dX^2_3+dX^2_4+dX^2_5,
\end{equation}
\begin{eqnarray}
 X_0 &=& r\sinh(t/\ell)\nonumber\\
 X_1&=&\sqrt{\ell^2+r^2},\nonumber \\
 X_2 &=& r\cosh(t/\ell)\cos\chi\nonumber\\
 X_3&=&r\cosh(t/\ell)\sin\chi\cos\theta,\nonumber \\
 X_4 &=& r\cosh(t/\ell)\sin\chi\sin\theta\cos\phi\nonumber\\
  X_5&=&r\cosh(t/\ell)\sin\chi\sin\theta\sin\phi.
\end{eqnarray}
Since $X_1\ge \ell$ in this embedding, the metric (\ref{global})  covers approximately only half of $AdS_5$.
One can check that the bifurcate hyperbolic horizon corresponds to $X_0=0$, $X_2=0$ and $0\leq r\leq \rho_{\rm{max}}$. To see this clearly, we present flat embedding of hyperbolic foliation of $AdS_5$
\begin{eqnarray}
  X_0 &=&\sqrt{\tilde{r}^2-\ell^2}\sinh(\tilde{t}/\ell),\nonumber \\
  X_1 &=& \tilde{r}\sqrt{1+\rho^2/\ell^2},\nonumber\\
  X_2&=&\sqrt{\tilde{r}^2-\ell^2}\cosh(\tilde{t}/\ell),\nonumber\\
  X_3 &=&  (\tilde{r}/\ell)\rho\cos\tilde{\theta},\nonumber\\
  X_4 &=& (\tilde{r}/\ell)\rho\sin\tilde{\theta}\cos\tilde{\phi}, \nonumber\\
  X_5 &=& (\tilde{r}/\ell)\rho\sin\tilde{\theta}\sin\tilde{\phi},
\end{eqnarray}
where we use letters with tilde to denote coordinates of the  hyperbolic foliation of $AdS_5$.
Thus horizon of hyperbolic foliation of $AdS_5$ is at $X_0^2=X_2^2$. The finite hyperbolic time cross section of hyperbolic
horizon resides at $X_0=0$ and $X_2=0$. In terms of global de Sitter coordinates,
it is $t=0$ and $\chi=\pi/2$. At $t$=0, the boundary theory ($r= \rho_{\rm{max}}$) is defined on $S^3$ and $\chi=\pi/2$ is
the equator dividing $S^3$ into two halves. Meanwhile, it can be found that the equator defined at $t=0$ and $\chi=\pi/2$ is
the event horizon to the north pole observer \cite{Witten:2001kn}. Taking the equator as the boundary of field theory defined on
one of the two halves. The holographic entropy is then \cite{Ryu:2006bv}
\begin{equation}\label{holoentropy}
    S^{hol}=\frac{Area(\Sigma)}{4G_5},
\end{equation}
where $\Sigma$ is a three dimensional minimal surface in $AdS_5$
bounded by the equator.

Although the half-global patch (\ref{global}) is time-dependent, the
equator on the boundary we are interested in sits at time $t=0$ thus we
expect that the minimal surface bounded by this equator also sits at $t=0$ due to the reflection
symmetry of (\ref{global}) in time ($t\rightarrow -t$).
 The surface $\Sigma$ has the area
\begin{equation}\label{area}
   Area(\Sigma)=4\pi\int^{\rho_{\rm{max}}}_0r^2\sin^2(\chi)\sqrt{r^2\chi'^2+\frac{1}{1+r^2/\ell^2}}dr,
\end{equation}
where $\chi'=\partial_r\chi$. The minimal surface solves the following equation
\begin{equation}
    r^2\sin(2\chi)\sqrt{r^2\chi'^2+\frac{1}{1+r^2/\ell^2}}=\partial_r\biggl(r^4\sin^2(\chi)\frac{\chi'}{\sqrt{r^2\chi'^2+\frac{1}{1+r^2/\ell^2}}}\biggr).
\end{equation}
A solution is just given by $\chi=\pi/2$. Inserting this into (\ref{area}), immediately we recover the result in (\ref{entropy})

\section{Holographic Cosmological Solutions Asymptotic to de Sitter}
We first note that the expanding de Sitter can be embedded in $AdS_5$ as
 \begin{equation}\label{FRW}
    ds^2_5=(1+\frac{r^2}{\ell^2})^{-1}dr^2+\frac{r^2}{\ell^2}(-dt^2+e^{2t/\ell}(dx^2+dy^2+dz^2)).
\end{equation}
In this section, we are interested in generalizing the above embedding to a more general and realistic cosmology.
For a generic scale factor $a^2(t)$, the nonvanishing components of Einstein tensor are obtained in the following
\begin{eqnarray}\label{ET}
  G_{rr} &=& \frac{6}{\ell^2+r^2}-\frac{3\ell^2}{r^2(1+r^2/\ell^2)}(\dot{H}+2(H^2-\ell^{-2})), \nonumber\\
  G_{tt} &=& -6r^2/\ell^4+3(H^2-\ell^{-2}),\nonumber\\
  G_{ij} &=& \delta_{ij}\biggl(6r^2a^2/\ell^4-(2\dot{H}+3(H^2-\ell^{-2}))a^2\biggr),
\end{eqnarray}
We require the following Einstein equations be satisfied
\begin{equation}\label{Eeom}
    G_{MN}=6\ell^{-2}g_{MN}+8\pi G_5T_{MN}\,,\qquad M, N=(r,t,x,y,z)
\end{equation}
where the energy momentum tensor is assumed to be
\begin{equation}
    T^{M}_{~N}=\frac{\ell^2}{r^2}\mbox{diag}\{q,-\rho,~p,~p,~p\}.
\end{equation}
The energy momentum conservation demands that
\begin{eqnarray}\label{emc}
  \dot{\rho}+3H(p+\rho) &=& 0, \nonumber\\
 r\partial_r q+2q+(\rho-3p) &=& 0.
\end{eqnarray}
We find that (\ref{Eeom}) and (\ref{emc}) are solved by
\begin{eqnarray}
&&  H^2 = \ell^{-2}+\frac{8\pi G_5}{3}\rho(t), \nonumber\\
 && \dot{\rho(t)}+3H\biggl(p(t)+\rho(t)\biggr) = 0,\\
 && q(t) = -\ft{1}{2} (\rho-3p).
\end{eqnarray}

In the following, we will discuss the constraints on the  state parameter $w$ from five dimensional energy conditions.
For $p=w\rho$, $q=-\ft{1}{2}(1-3w)\rho$
\begin{itemize}
  \item 1) Weak energy condition $\rho\geq0, \rho+p\geq0$ and $\rho+q\geq0$ require $w\geq-\ft{1}{3}$,
  \item 2) Null energy condition $\rho+p\geq0$ and $\rho+q\geq0$ gives the same answer if $\rho\geq0$,
  \item 3) Strong energy condition $\rho+3p+q\geq0$, $\rho+p\geq0$ and $\rho+q\geq0$ demands that $w\geq-\ft{1}{9}$,
  \item 4) Dominant energy condition $\rho\geq0$, $\rho\geq|p|$ and $\rho\geq|q|$ reads $-\ft{1}{3}\leq w\leq1$.
\end{itemize}

To end this section, we should mention that the five dimensional embedding of inflationary universe (except the
pure de Sitter case) has a naked singularity at $r=0$ which can be seen below. The curvature square is
\begin{equation}
    R_{MNPQ}R^{MNPQ}=16\ell^{-4}+12(\ell^{-2}-\frac{\ell^2H^2-1}{r^2})^2+12(\ell^{-2}-\frac{\ell^{2}\dot{H}+\ell^2H^2-1}{r^2})^2.
\end{equation}
Since near $r=0$, the $(t,r)$ component of the five dimensional embedding of an inflationary universe with matter
 approaches the Rindler metric, the singularity at $r=0$ lies on the Rindler horizon and is null. We also made attempts to hide this singularity behind a black hole horizon but it turns out that new singularity will appear at the horizon, thus there seems to be always a null singularity in the simple embedding of time dependent solutions.

\section{ Conlusion}
In this paper, we try to understand quantum field theory in de Sitter via the AdS/CFT correspondence. We find that the
static de Sitter is conformally related to $R\times H^3$, and the latter can be embedded into the hyperbolic $AdS_5$ as
conformal boundary. The hyperbolic $AdS_5$ has a horizon thus has a temperature. The temperature is precisely the de Sitter
 temperature. We compute energy and entropy of the hyperbolic $AdS_5$ which correspond to the same quantities of
 certain quantum field theory living on the boundary. It turns out that energy is proportional to the size of the
  dS horizon and entropy is proportional to the horizon area if an IR cut-off is introduced in the bulk.
  The functional dependence of energy and entropy with respect to horizon size recovers previous results using quantum field theory method.

We also tackle the problem of embedding inflationary dS into $AdS_5$. It is found that the solution
can be obtained by introducing a simple matter energy momentum tensor in the bulk. An imperfect point
is that this embedding often has naked null singularity which can not be resolved in a simple way. Thus how to deal with this singularity deserves further study.

\section*{Acknowledgement}
This work is supported by the NSFC under Grant Nos.10535060/A050207, 10975172 and 10821504, and Ministry of Science and Technology 973 program under
grant No.2007CB815401.

\end{document}